\documentclass[iop]{emulateapj}
\usepackage{apjfonts}

\begin{document}


\title{On the Formation of Ultraluminous X-ray Sources with Neutron Star Accretors: \\ the Case of M82 X-2}
\author{Tassos Fragos$^1$, Tim Linden$^{2}$, Vicky Kalogera$^{3}$, and Panos Sklias$^1$}
\affil{$^1$ Geneva Observatory, University of Geneva, Chemin des Maillettes 51, 1290 Sauverny, Switzerland}
\affil{$^2$ The Kavli Institute for Cosmological Physics, University of Chicago, Chicago, IL 60637, USA}
\affil{$^3$ Center for Interdisciplinary Exploration and Research in Astrophysics (CIERA) \& Dept. of Physics and Astronomy, Northwestern University, 2145 Sheridan Rd, Evanston, IL 60208, USA;}
\shortauthors{Fragos et al.}
\shorttitle{On the Formation of NS ULXs: the Case of M82 X-2}
\keywords{stars: binaries --- stars: neutron ---  X-rays: binaries}

\begin{abstract}
The recent discovery of a neutron star accretor in the ultra-luminous X-ray source M82 X-2 challenges our understanding of high-mass X-ray binary formation and evolution. By combining binary population synthesis and detailed mass-transfer models, however, we show that the binary parameters of M82 X-2 are not surprising provided non-conservative mass transfer is allowed. Specifically, the donor-mass lower limit and orbital period measured for M82 X-2 lie near the most probable values predicted by population synthesis models, and systems such as M82 X-2 should exist in approximately 13\% of the galaxies with a star-formation history similar to M82. We conclude that the binary system that formed M82 X-2 is most likely less than 50\,Myr old and contains a donor star which had an initial mass of approximately 8-10 M$_\odot$, while the NS's progenitor star had an initial mass in the $8-25\,\rm M_{\odot}$ range. The donor star still currently resides on the main sequence, and is capable of continued MT on the thermal timescale, while in the ultra-luminous X-ray regime, for as long as 400,000 years.
\end{abstract}

\section{Introduction}
\label{sec:introduction}

Ultra-luminous X-ray Sources (ULX) are among the most extreme phases of binary  evolution, characterized by X-ray luminosities exceeding 10$^{39}$~erg~s$^{-1}$~\citep{2009MNRAS.397.1836G}. Since these systems exceed the Eddington luminosity for compact objects formed via stellar evolution, models must invoke one of two mechanisms in order to account for their high luminosities. Under the assumption of spherical, Eddington-limited accretion, one class of models employs intermediate-mass black holes (BHs) with masses exceeding $\sim$\,100~M$_\odot$~\citep{1999ApJ...519...89C, 2004cbhg.symp...37V, 2008ApJ...688.1235M, 2011NewAR..55..166F}. The second method employs some combination of thin accretion disks \citep{1973A&A....24..337S,2002ApJ...568L..97B}, whose luminosity can exceed the Eddington limit without being disrupted by radiation pressure, and/or anisotropic X-Ray emission \citep{2009MNRAS.393L..41K}, in order to reproduce the luminosity of observed systems without exceeding the Eddington limit. The formation rate of ULXs has been accounted for in models examining the formation rate of the most massive BHs from single stars~\citep[e.g.][]{2010MNRAS.408..234M} and with the incorporation of binary evolution effects~\citep[e.g.][]{2010ApJ...725.1984L}. 

Recently, NuSTAR discovered a pulsar spatially coincident with the location of the ULX M82 X-2, indicating that the compact object in this system is a neutron star \citep[NS;][]{2014Natur.514..202B}. Assuming a typical NS mass  $M_{\rm NS}\sim 1.4\,\rm M_\odot$, the measured mass function and orbital period (2.52\,days) indicate that the donor star has a mass $M_2 \gtrsim 5.2\,\rm M_\odot$ and a radius $R_2\gtrsim 7\,\rm R_\odot$. The X-ray luminosity of M82 X-2 is observed to be L$_X$(0.5 -- 10 keV)~=~6.6~$\times$~10$^{39}$~erg~s$^{-1}$, indicating that the system has an accretion rate that exceeds the Eddington limit by at least a factor of 30.

The identification of M82 X-2 as a NS ULX has already generated a variety of new theoretical insights. \citet{2014arXiv1411.3168T} finds that the spin-up behavior of the pulsar is consistent with a low-magnetic field ($\sim1\,\rm TG$) magnetar, while \citet{2014arXiv1410.5205E} and \citet{2014arXiv1412.1823D} use the physical properties of the system to imply larger magnetic fields $\sim$10-100~TG. \citet{2015MNRAS.448L..43K} note that the high MT rates would spin up the NS to millisecond periods within 10$^5$~yr, opening the possibility that observable high-mass X-ray binaries may be found with a millisecond pulsar accretor. \citet{2014arXiv1410.8745L} suggests that the extremely super-Eddington accretion rates require a new MT regime, where an optically thick accretion curtain shields the interior gas from the outgoing X-ray flux. Furthermore, this analysis suggests that the equilibrium, where the Alfv\'en radius matches the co-rotation radius, indicates a magnetic field of 14~TG. An analysis by \citet{2014arXiv1411.5434C} disagrees, arguing that the luminosity of M82 X-2 can be explained just with mild geometrical beaming. 

Apart from the specific emission mechanism and the magnitude and influence of the NS's magnetic field, the measured orbital period and the lower limits on the donor mass and radius are both telling and puzzling at first glance: (1) the high inferred accretion rate onto the NS requires that the donor is in Roche-lobe overflow (RLO), as wind-fed X-ray binaries with NS accretors are expected to have X-ray luminosities orders of magnitude below the ULX range~\citep{2010ApJ...725.1984L}, (2) the donor star must be hydrogen rich, as a helium star with mass  $\gtrsim 5\,\rm M_\odot$ cannot fill its Roche lobe in a $2.5\,\rm day$ period orbit, independent of its evolutionary stage, (3) the evolutionary mechanism must have either a long lifetime, or a high formation rate, in order for this one system to exist in the local universe. 

Still, the picture of a donor more massive than $\simeq\,5\,\rm M_\odot$ transferring mass onto a NS is particularly puzzling. While RLO X-ray binaries with NS accretors are a common output of population synthesis models, they typically involve lower-mass donors \citep[e.g.][]{Fragos2008,Fragos2009,Fragos2013,Fragos2013b}. This is due, primarily, to  the stability criteria for RLO from a massive donor star onto a lower-mass accretor. Early polytropic and semi-analytic models, with the assumption of conservative MT, demonstrated that systems with a donor more than $\sim$3~times as massive as the compact object would quickly enter a delayed dynamical instability~\citep{1987ApJ...318..794H, 2004ApJ...601.1058I}, producing a stellar merger. However, more recent studies of the stability of binary MT phases, using detailed binary evolution codes, showed that some of the approximations made in earlier studies, such as fully adiabatic mass-loss and the strict enforcement of hydrostatic equilibrium, underestimate the maximum mass-ratio of a binary in which dynamically stable MT can occur \citep[e.g.][]{2013A&ARv..21...59I,Pavlovskii:2014uk}. A crucial factor in determining this maximum mass ratio for stability of non-conservative MT phases is the specific angular momentum of the material ejected from the binary. Several different models have been developed for different physically motivated mass-loss models \citep[e.g.][]{1994inbi.conf..263V, 1999MNRAS.309..253K, TvdH2006}.

In this \emph{Letter}, we investigate the formation of ULX systems with NS accretors and derive constraints on the evolutionary history and current properties of M82 X-2. We first utilize the Binary Stellar Evolution (BSE) code~\citep{Hurley2002} in order to study the initial binary properties that can produce a MT event with characteristics similar to M82 X-2. We terminate simulations with this code at the onset of RLO and perform the MT calculations with the detailed stellar evolution code MESA \citep{Paxton2011,Paxton2013}, in order to accurately predict the duration and accretion rate of the MT phase. 

\section{The host galaxy: M82}
\label{sec:M82}

M82 is one of the nearest dusty starburst galaxies. It has sustained vigorous star formation in the past 50 to 100 Myr \citep{2008A&A...484..711B}, probably triggered by its anterior interaction with M81. Based on publicly available photometry from \emph{IRAS}, \emph{Spitzer}-MIPS, and \emph{Herschel}-SPIRE \citep{2010A&A...518L..66R}, we estimate the IR-derived star-formation rate (SFR) to be of at least $5.5\,\rm M_\odot\,yr^{-1}$ during this period, in agreement with \citet{2003PASP..115..928K}. We should note, however, that preliminary spectral energy distribution fitting results, following the methodology described in \citet{2014A&A...561A.149S}, favor a declining star-formation activity in M82 over a constant SFR. Along the same lines, earlier, spatially resolved, studies infer SFRs as high as $\sim 30\,\rm M_\odot\,yr^{-1}$, depending on various IMF assumptions \citep[e.g.][]{2003ApJ...599..193F}. In the rest of this paper, we adopt a constant SFR of $5\,\rm M_\odot\,yr^{-1}$, over the last 100\,Myr, and a metallicity of $Z=0.02$ \citep{2001ApJ...552..544F} for all our simulations.

\section{Numerical Models}
\label{sec:numerical}
\subsection{Binary Population Synthesis with BSE}
\label{sec:bse}

In order to determine the most probable initial parameters for M82 X-2, we calculated the evolution of $10^7$ isolated\footnote{Formation channels through dynamical interactions in dense stellar systems are not taken into account} massive binaries, restricting our modeling to systems with a primary mass large enough to form a NS. We utilize the BSE population synthesis code \citep{Hurley2002}, modified to include the suite of stellar wind prescriptions for massive stars described in \citet{Belczynski2010}, and the fitting formulae for the binding energy of the envelopes of stars derived by \citet{2011ApJ...743...49L}. We followed binary systems through any binary interaction prior to the primary's supernova, through the supernova explosion which formed the NS, and the post-supernova detached evolution. We halted the code at the moment when the secondary star fills its Roche lobe, and recorded the state of the binary to use as an input to our stellar evolution and MT code.

In our modeling, we assume a \citet{Kroupa2001} initial mass-function, a flat distribution of initial binary mass-ratios, a logarithmically flat distribution of initial orbital separations, a Maxwellian distribution of NS supernova kick magnitudes with $\sigma=265\,\rm km\,s^{-1}$ \citep{Hobbs2005}, and a common envelope efficiency of $\alpha_{CE}=0.5$. These parameter values have been shown to produce synthetic populations of X-ray binaries that are in good agreement with observations of X-ray luminosity functions and X-ray scaling relations of extragalactic populations. \citep[e.g.][]{Fragos2008,Linden2009,Fragos2013,2013ApJ...774..136T}. We repeated all calculations in this paper utilizing common envelope efficiencies of $\alpha_{CE}=0.2$ and $\alpha_{CE}=1.0$, finding only a small change in the resulting distributions for the orbital periods and donor masses of possible M82 X-2 progenitors. The qualitative insensitivity of our models to the common envelope efficiency is explained by the high mass of the secondary star ($\gtrsim 5\,\rm M_{\odot}$) which guaranties that the orbital energy of the binary is usually greater than the binding energy of the NS's progenitor envelope. 



\begin{figure*}
\plottwo{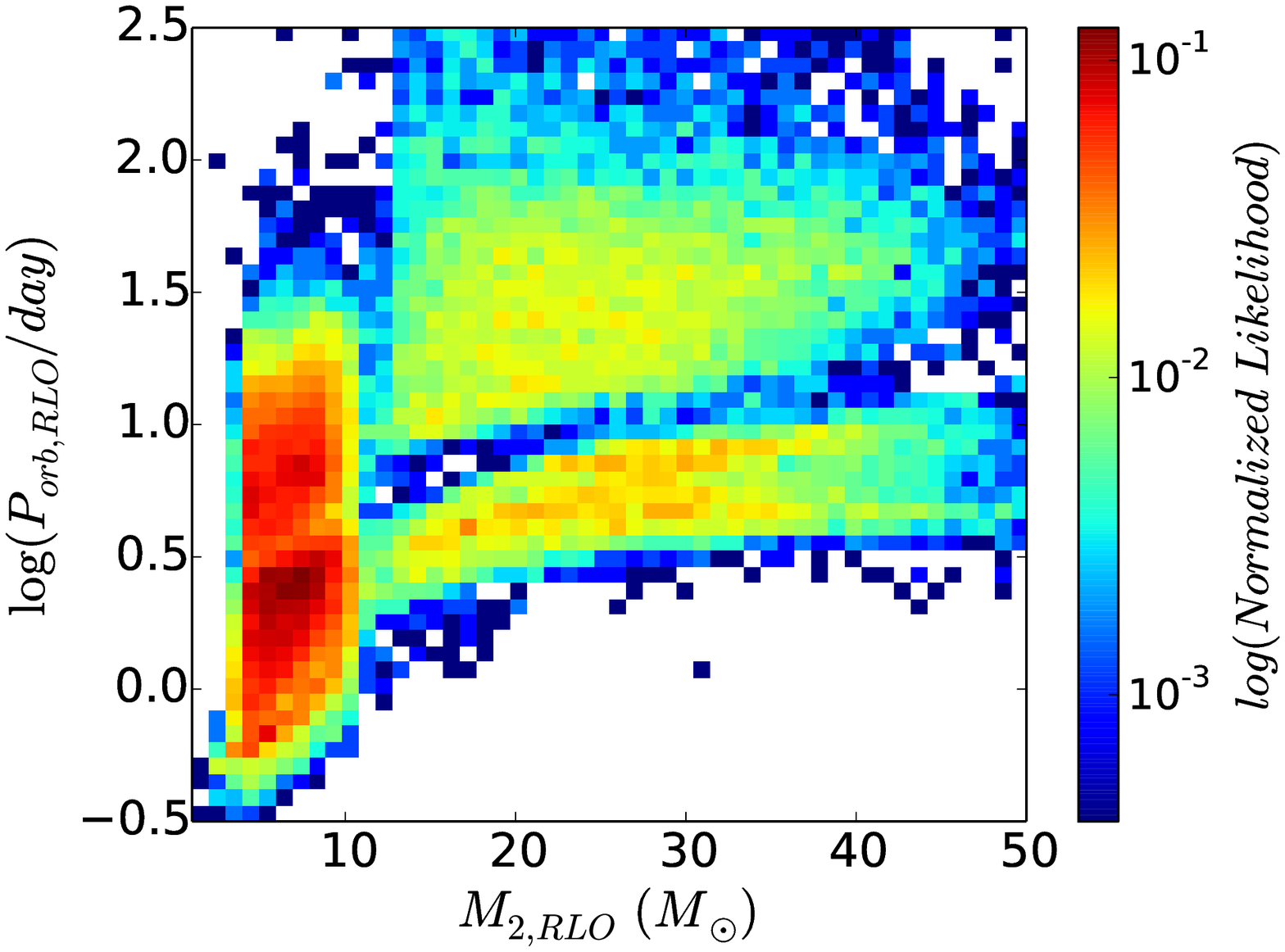}{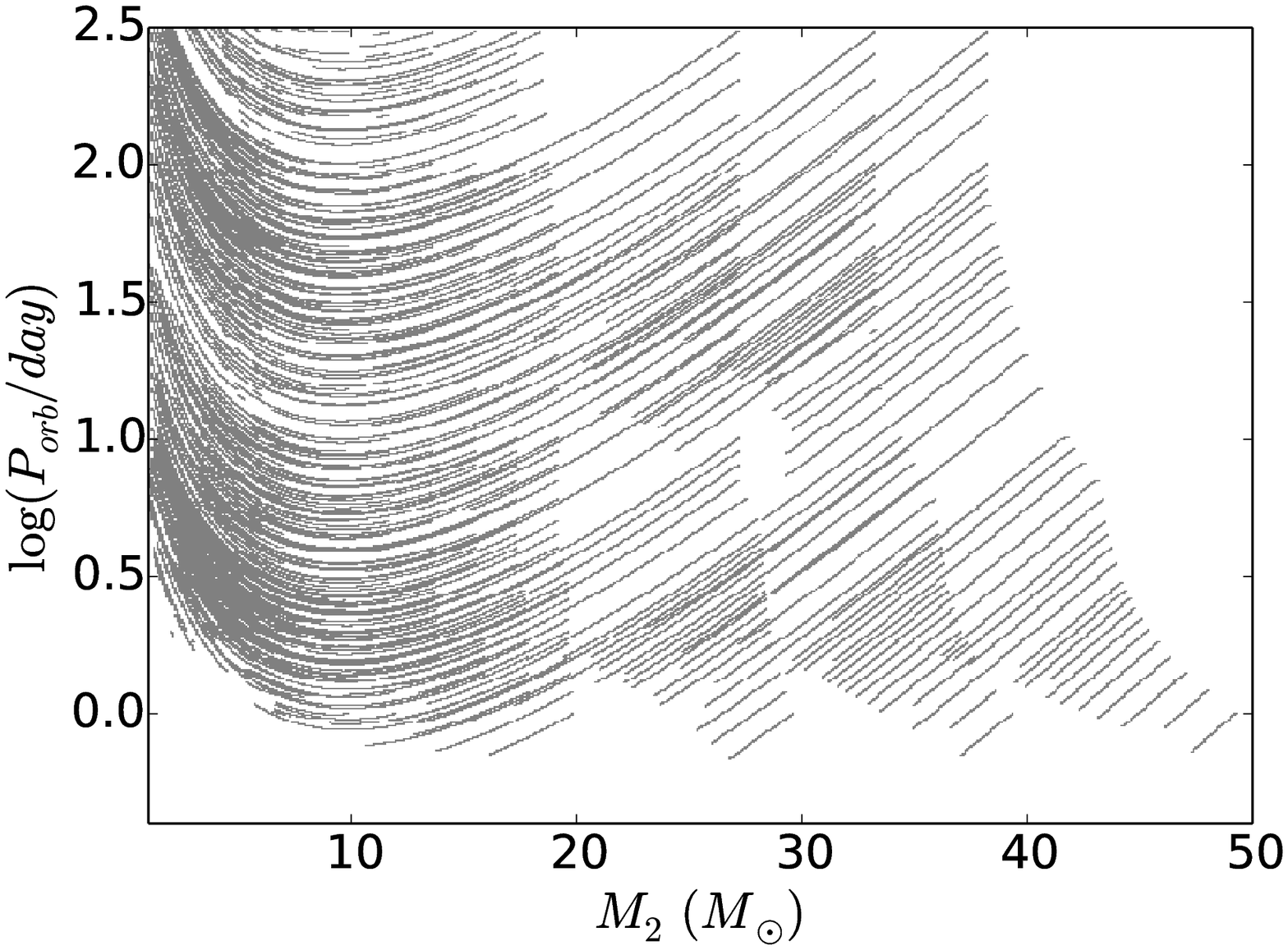}

\caption{\textbf{Left panel:} The two-dimensional distribution of binary orbital periods versus donor masses of binaries that contain a NS and a secondary non-degenerate star that just filled its Roche lobe, calculated from $10^7$ isolated binaries evolved with BSE. 
\textbf{Right panel:} The gray shaded area shows the parameter space in the orbital period-donor star mass plane  that is covered by our grid of detailed MT sequences calculated with MESA, assuming 
a MT efficiency parameter $\alpha=0.9$. 
\label{fig:bse_distributions}}
\end{figure*}

\subsection{Detailed Mass-Transfer Calculations with MESA}
\label{sec:mesa}

We use version 7184 of the stellar evolution code MESA \citep{Paxton2011, Paxton2013} in order to calculate grids of $\sim 2,700$ MT sequences for NS X-ray binaries undergoing MT. We assume an initial NS mass of $1.4\,\rm M_{\odot}$, and cover the parameter space of possible orbital periods and companion masses at the onset of RLO. Specifically we consider initial donor masses ($M_{2}$) between $1.0\,\rm M_\odot$ and $50.0\,\rm M_\odot$
 and initial binary orbital periods between $1.0$ and $300.0\,\rm day$.
 A higher initial NS mass (e.g. $2.0\,\rm M_{\odot}$), as suggested by \citet{2011MNRAS.416.2130T} for the formation of the binary millisecond pulsar PSR J1614-2230, will yield in our case qualitatively identical results, and is not considered in this work as it would increase significantly the computational cost of our simulations.

For our MT calculations we employ the implicit MT scheme and the hydrodynamic solver of MESA, in order to properly resolve the onset of a dynamically unstable MT phase. We terminate the MT sequences when any of the following criteria are met: \emph{(i)} the mass of the donor star drops below $1\,\rm M_{\odot}$, \emph{(ii)} the age of the system exceeds the age of the Universe ($13.7\,\rm Gyr$), \emph{(iii)} the whole envelope of the donor star is removed, or \emph{(iv)} the timestep of the hydrodynamic solver drops below $10^{-6}\,\rm s$, while the MT rate exceeds $10^{-2}\,\rm M_{\odot}\,yr^{-1}$, and no solution can be found such that the radius of the donor is less or equal to the radius of its Roche lobe. We consider the latter as the onset of a common envelope.

The secular evolution of the orbit of a mass-transferring binary depends on the accretion efficiency, and on the specific angular momentum that is lost from the system. Here, following \citet{1994inbi.conf..263V}, we define a parameter $\alpha$ denoting the fraction of the mass lost from the donor ($\dot{M}_2$) that is ejected from the vicinity of the donor star in the form of ``fast wind'', i.e. with specific angular momentum equal to that of the donor star. The remainder of the mass ($(1-\alpha)\dot{M}_2$) is funneled through the first Lagrangian point to the accretion disk around the NS. We limit the accretion rate onto the NS to 100 times the Eddington limit ($\dot{M}_{Edd}$\footnote{For the calculation of $\dot{M}_{Edd}$ a radiative efficiency of 10\% is assumed.}). We note here that the exact value of this upper limit, within a factor of a few, does not play a significant role in the evolution of a mass-transfer sequence. The fraction of $(1-\alpha)\dot{M}_2$ that is in excess of $100\dot{M}_{Edd}$ is again ejected from the system in the form of ``fast winds''. However,
we assume that this mass is ejected from the vicinity of the NS and has the specific angular momentum of the accretor.  Since these X-ray binaries have a donor star that is more massive than the NS, mass lost from the vicinity of the donor star has a smaller angular momentum than mass lost from the vicinity of the accretor.  Hence, a value $\alpha > 0$ would results in a reduced rate of angular momentum loss from the orbit, and would therefore increase the stability of the MT phase. We ran our grid of MT sequences for 4 different values of the parameter $\alpha$ ($\alpha = 0.0,\, 0.5,\, 0.9,\, 0.99$) in order to study the effect that it has on the formation of potential NS ULXs. The right panel of Figure~\ref{fig:bse_distributions} shows the parameter space in the orbital period-donor star's mass  that is covered by our grid of detailed MT sequences, assuming an initial NS mass of $1.4\,\rm M_\odot$ and a MT efficiency parameter $\alpha=0.9$

\section{Results}
\label{sec:results}

In Figure~\ref{fig:bse_distributions} (left) we show the distribution of orbital periods and donor masses, at the beginning of the MT phase, as produced by the BSE code. Intriguingly, we note that the peak of this distribution is centered around systems with a donor mass of 5-10~M$_\odot$ and an orbital period between 1.6 -- 3.2~days at the onset of RLO. This distribution of binaries is thus highly compatible with the observation of M82 X-2. The population appears to change in character at $\simeq$13\,M$_\odot$; systems to the left of this boundary experience a CE phase before NS formation, while systems to the left evolve through stable MT between the two initial massive stars. Similarly, there is a ``valley'' around orbital periods of 3-10\,days, which separates systems with main-sequence donors from systems with evolved donors. However, in order to estimate the probability that any of these systems is currently a ULX, we need to convolve this distribution function of binaries at the onset of RLO with the detailed MESA calculations, so that we determine the duration of ULX activity in each binary. For our analysis using the MESA code, we define as ULX any X-ray binary in which the NS is accreting material at a rate that exceeds $10\dot{M}_{Edd}$.


\begin{figure*}
\plotone{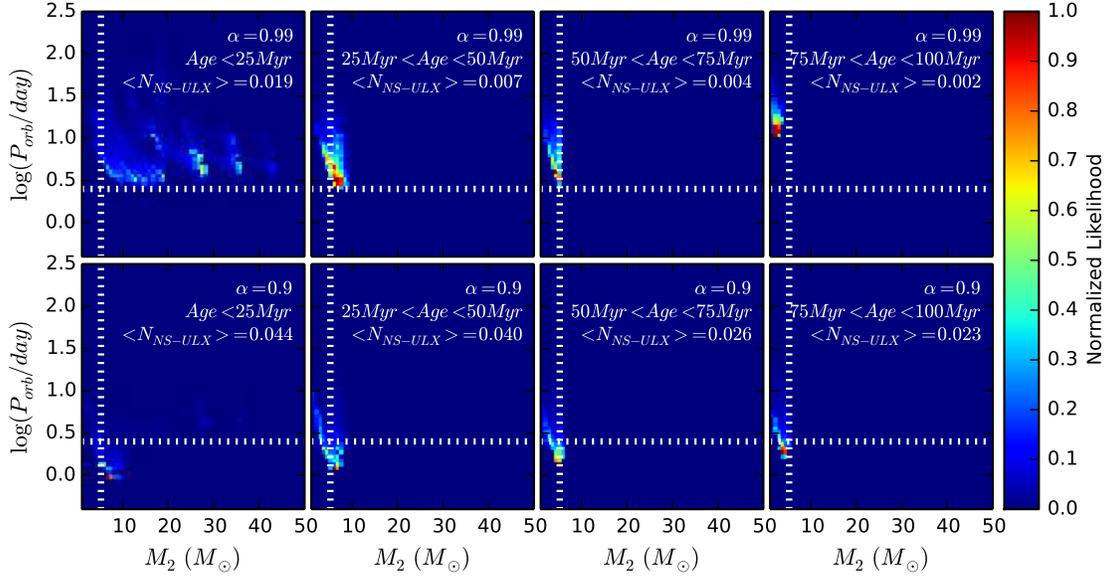}

\caption{Relative likelihood of a NS ULX existing \emph{today} at a given orbital period and donor mass. Each column of the figure corresponds to a different age range of the ULXs, while each row corresponds to a different choice of the accretion parameter $\alpha$ (top: $\alpha=0.99$, bottom: $\alpha=0.9$). \label{fig:mesa_likelihood1}}
\end{figure*}

\begin{figure*}
\plotone{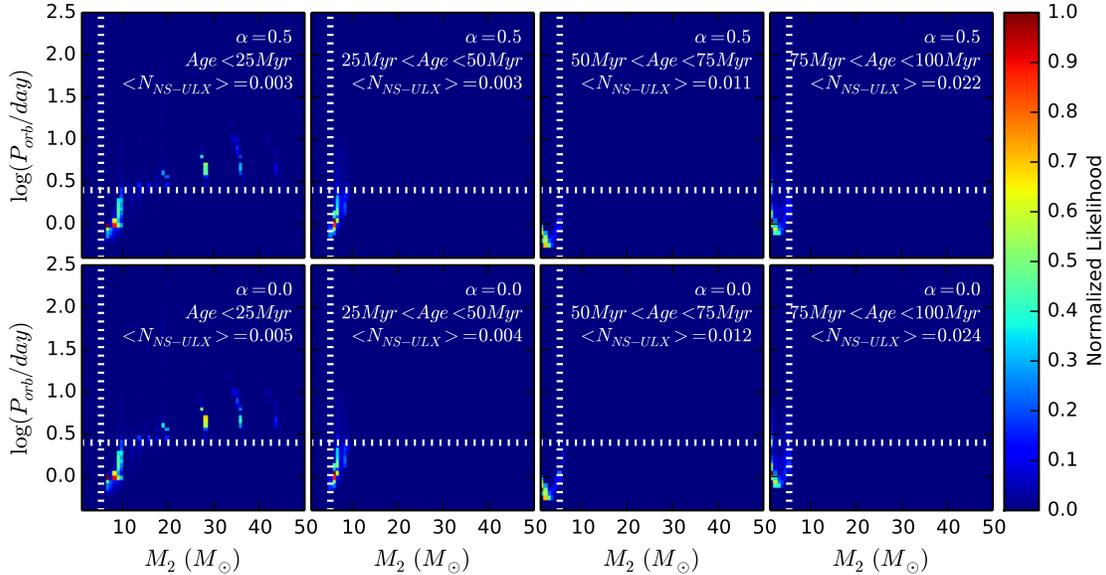}

\caption{Same as Figure~\ref{fig:mesa_likelihood1}, but for $\alpha=0.5$ (top) and $\alpha=0.0$ (bottom). \label{fig:mesa_likelihood2}}
\end{figure*}

In Figures~\ref{fig:mesa_likelihood1} and \ref{fig:mesa_likelihood2} we show the resulting probability distribution of having a ULX with a NS accretor in the parameter space of the observed characteristics of M82 X-2 (donor mass and orbital period). We show results for four different age ranges for the ULX, 0~--~25~Myr, 25~--~50~Myr, 50~--~75~Myr, and 75~--~100~Myr, and four different values of non-conservative MT: $\alpha~=~0.99,\,0.9,\,0.5\ \rm and\ 0.0$. The color bar depicts the relative probability of a NS ULX existing  today with the given orbital parameters, and the expected number of  systems in M82 is listed as $<N_{\rm NS-ULX}>$. This probability is calculated by assigning to each one of our MT sequences a weight proportional to the formation rate of NS binaries reaching RLO at periods and donor masses close to that of the MT sequence, as calculated by our BSE population synthesis models. In addition, we weight each MT sequence by the time that the sequence spends at each period and donor mass, while the MT rate exceeds $10\dot{M}_{Edd}$. 

In each panel of Figures~\ref{fig:mesa_likelihood1} and \ref{fig:mesa_likelihood2}, we show white lines denoting the observed orbital period and minimum donor mass of M82 X-2. We note two key observations: (1) the observed parameters of M82 X-2 are well matched to the most probable parameters of a ULX with a NS accretor, (2) the probability of such a system in M82 existing is relatively high, between $\sim 0.03$ and $\sim0.13$ (i.e. we expect to have a NS ULX in every one out of approximately 8 galaxies, with properties similar to M82). Furthermore, comparing the observed properties of M82 X-2 to the different panels of Figures~\ref{fig:mesa_likelihood1} and \ref{fig:mesa_likelihood2}, we can put limits on both the current age of the system and the accretion efficiency of the MT phase. Accretion efficiencies below 1\% ($\alpha>0.99$) can be excluded, as the accretion rate becomes too low to produce a ULX (see the very low expected number of ULX in the all panels of the first row of Figure~\ref{fig:mesa_likelihood1}). Similarly, accretion efficiencies above 50\% ($\alpha<0.5$) predict that ULXs would be formed at significantly shorter periods than observed in M82 X-2. The age of the population plays also role, as older systems tend to have shorted periods and lower mass companions. Based on this comparison, a system like M82 X-2, would likely have an age $\lesssim75\rm\,Myr$, and an accretion efficiency below 10\%.


\begin{figure*}
\plotone{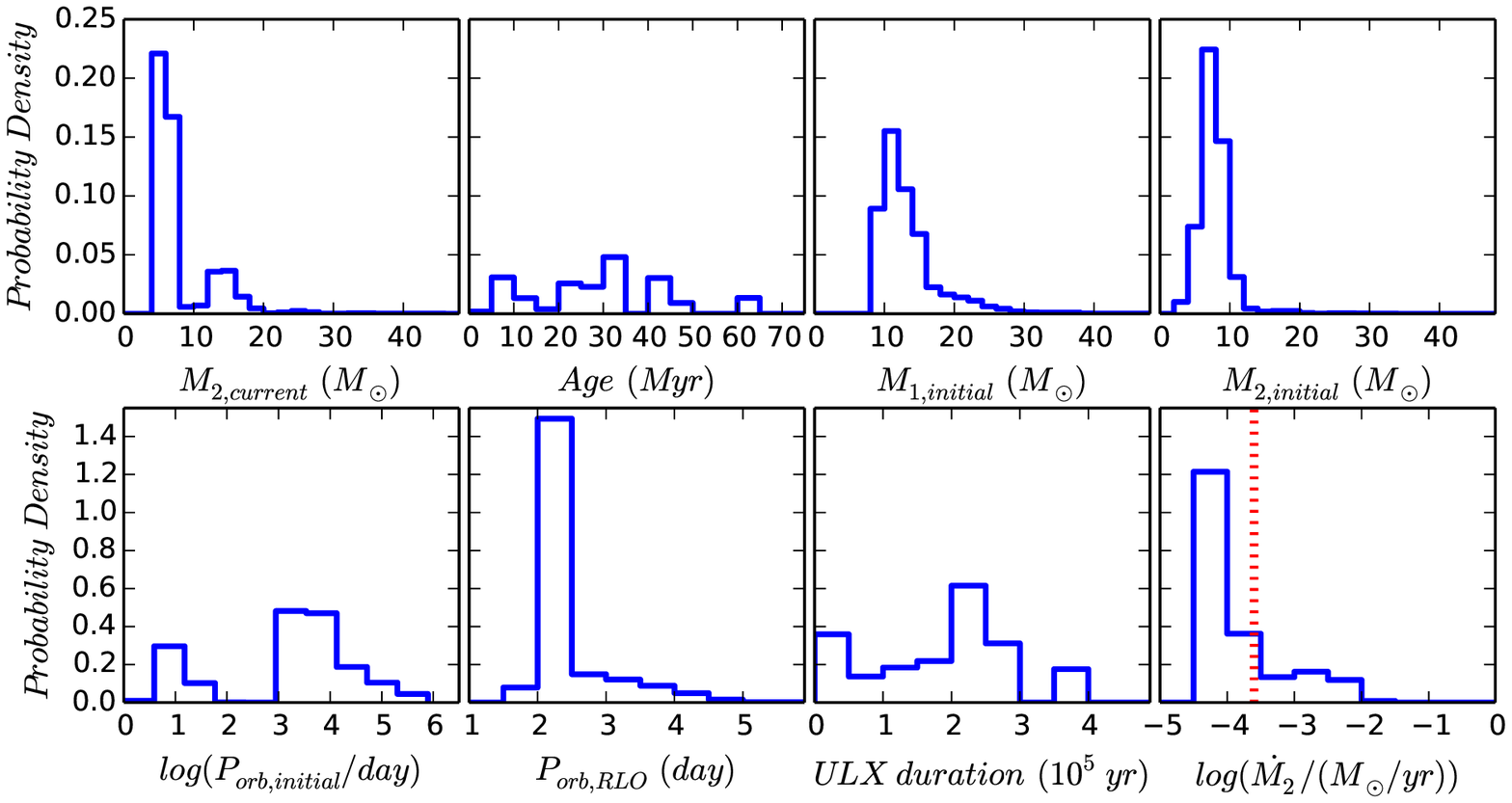}

\caption{Probability density functions of the current mass and age of the companion star in M82 X-2, the initial component masses and orbital period of M82 X-2's progenitor binary, the orbital period of the binary at the onset of RLO, as well as the duration of the ULX phase and the current mass-loss rate of the companion star. We take into account observational selection effects by weighting all probability density functions with the total time that each modeled system has properties similar to M82 X-2, i.e. $M_2>5.2\,M_{\odot}$, $\dot{M}_{\rm accreted}>10\dot{M}_{\rm Edd}$, and $2<P_{orb}<3\,day$. Finally, in the bottom right panel we show for comparison the MT rate corresponding to the Kelvin-Helmholtz thermal timescale of an $8\,M_{\odot}$ star that is just filling its Roche lobe in a $2.5\,day$ orbit around a $8\,M_{\odot}$ NS. \label{fig:histograms}}
\end{figure*}

We can derive more robust constraints on both the current properties of M82 X-2 and the properties of its progenitor binary by following an analysis similar to \citet{Fragos2009b} and \citet{2015ApJ...800...17F}. For each MT sequence in our grid, we examine all systems that simultaneously fit the observational constraints on M82 X-2, specifically an orbital period of  2.52\,day, a donor mass above 5.2\,$\rm M_\odot$ and an accretion rate onto the NS higher than $10\dot{M}_{Edd}$. In this part of the analysis we only consider one of our MT grids, with $\alpha=0.9$, which predicts the highest expected number of NS ULXs and the closest-matching properties for M82 X-2. In Figure~\ref{fig:histograms} we show the probability density functions of current binary parameters (donor mass and age, and ULX duration) for systems with the M82 X-2 measured characteristics. We also use the results from our parameter-space exploration in BSE in order to derive self-consistent  constraints on the initial binary parameters of M82 X-2's progenitor (component masses and orbital period).

\section{Discussion}
\label{sec:conclusions}

By combining a parameter space exploration using the BSE population synthesis code with a detailed treatment of MT using the MESA code, we have shown that the properties of M82 X-2 are well explained by current models of binary population synthesis. Specifically, our models reveal the following details about the formation of M82 X-2:

\begin{itemize}
\item{The donor star is hydrogen rich, as a helium star of mass $\gtrsim 5.2\,\rm M_{\odot}$ cannot fill the Roche lobe of a $\sim 2.5$~day orbit,  even at the late stages of the carbon/oxygen burning phase. Furthermore, the donor star is in RLO, as wind-fed NS binaries are expected to have X-ray luminosities below the Eddington limit \citep[e.g.][]{2010ApJ...725.1984L}. }

\item{The orbital period of any ULX with a NS accretor is most likely to be observed with an orbital period between 1--3 days, and a donor mass between 3--8~M$_\odot$. This places the observed properties of M82 X-2 near the peak of the likelihood distribution for this class of systems.}
   
\item{The MT is highly non-conservative and happens on the thermal timescale. Conservative MT leads to dynamical instability (possibly delayed) as discussed in the literature \citep[e.g.][]{1987ApJ...318..794H,2004ApJ...601.1058I}. The accretion efficiency must be $\lesssim 0.1$, consistent with local ULX analogues \citep[e.g.][]{Neilsen:2009bh,Ponti:2012ip}, and mass lost should be ejected from the system in the form of ``fast winds'' from the vicinity of the donor, i.e. it should carry relatively little angular momentum in order to provide stability. A good example of this is the short-period high-mass X-ray binary Cygnus X-3 \citep{1994inbi.conf..263V}.}
 
 \item{Assuming an accretion efficiency of 10\%, we estimate that the number of NS ULXs per unit of SFR is $N_{NS-ULX}/SFR = 0.027\,\rm M_{\odot}^{-1}\,yr$. This number is an order of magnitude lower compared to predictions in the literature for the formation rate of ULXs with BH accretors, which suggest a $N_{BH-ULX}/SFR \sim 0.2 - 1.0\,\rm M_{\odot}^{-1}\,yr$ at solar metallicity \citep{2005MNRAS.356..401R,2010ApJ...725.1984L}. We stress here that for a NS ULX to be identified as such, the  requirement of a highly magnetised NS, that allows the production of X-ray pulses, should be taken into account. The modelling of the NS's magnetic field is outside the scope of this work.}

\end{itemize}

\acknowledgments
TF acknowledges support from the Ambizione Fellowship of the Swiss National Science Foundation (grant PZ00P2\_148123). TL is supported by the National Aeronautics and Space Administration through Einstein Postdoctoral Fellowship Award Number PF3-140110. VK acknowledges support through NASA ADP grant NNX12AL39G. VK and TL acknowledge useful discussions with Maxim Lyutikov. The computations of this work were performed at University of Geneva on the Baobab cluster.


\end{document}